\documentclass[a4paper,11pt]{article}
\usepackage[colorlinks,citecolor=red,urlcolor=blue,bookmarks=false,hypertexnames=true]{hyperref} 

\usepackage{cite}

\usepackage{amssymb,amsmath,amsfonts}
\usepackage{float}
\usepackage{relsize}
\usepackage{graphicx}
\usepackage{tikz}
\usetikzlibrary{patterns}
\usetikzlibrary{shapes, arrows, shadows}

\tikzstyle{vertex} = [circle, draw, fill=blue!20, scale=1,auto=left]
\tikzstyle{vert} = [circle, draw, fill=blue!20, scale=.8,auto=left]
\tikzstyle{line} = [draw]
\newcommand{\midarrow}{\tikz \draw[-triangle 60] (0,0) -- +(.1,0);}

\usepackage{upgreek}

\usepackage{pgfplots}

\pgfplotsset{compat=newest}

\pgfdeclarepatternformonly[\LineSpace]{my north east lines}{\pgfqpoint{-1pt}{-1pt}}{\pgfqpoint{\LineSpace}{\LineSpace}}{\pgfqpoint{\LineSpace}{\LineSpace}}%
{
    \pgfsetlinewidth{0.4pt}
    \pgfpathmoveto{\pgfqpoint{0pt}{0pt}}
    \pgfpathlineto{\pgfqpoint{\LineSpace + 0.1pt}{\LineSpace + 0.1pt}}
    \pgfusepath{stroke}
}

\newdimen\LineSpace
\tikzset{
    line space/.code={\LineSpace=#1},
    line space=3pt
}

\def\be{\begin{equation}}
\def\ee{\end{equation}}
\def\bea{\begin{eqnarray}}
\def\eea{\end{eqnarray}}
\def\pt{\partial_{t}}

\begin{document}

\begin{titlepage}
\date{\today}       \hfill

\begin{center}

\vskip .2in
{\LARGE \bf   Ising model in a boundary magnetic field with random discontinuities}\\
\vspace{5mm}

\today
 
\vskip .250in

\vskip .3in
{\large Anatoly Konechny}

\vskip 0.5cm
{\it Department of Mathematics,  Heriot-Watt University\\
Edinburgh EH14 4AS, United Kingdom\\[10pt]
and \\[10pt]
Maxwell Institute for Mathematical Sciences\\
Edinburgh, United Kingdom\\[10pt]
}
E-mail: A.Konechny@hw.ac.uk
\end{center}

\vskip .5in
\begin{abstract} \large
We consider a two-dimensional Ising field theory on a space with boundary in the presence of a
piecewise constant boundary magnetic field which is allowed to change value discontinuously along the boundary. 
We assume zero magnetic field in the bulk. The positions of discontinuities are averaged over as in the annealed disorder. 
This model is described by a boundary field theory in which a superposition of the free spin boundary condition is perturbed   
by a collection of boundary condition changing operators. The corresponding boundary couplings give the allowed constant values 
of the magnetic field as well as the fugacities for the transitions between them.  
We show that when the value of the magnetic field is allowed to take only two different values which are the same in magnitude but have different signs  the model can be described by a quadratic Lagrangian. We calculate and analyse the exact reflection matrix for this model. We also calculate the boundary entropy and study in detail the space of RG flows in a three-parameter space and with four different infrared fixed points. We discuss the likely breakdown of integrability  in the extended model which allows for two generic values of the boundary magnetic field, backing it by some  calculations.

\end{abstract}

\end{titlepage}

\renewcommand{\thepage}{\arabic{page}}
\setcounter{page}{1}
\large 

\section{Introduction }
\renewcommand{\theequation}{\arabic{section}.\arabic{equation}}

The two-dimensional Ising model on a half-plane with a boundary magnetic field was considered in \cite{MW} and later in 
\cite{Bariev}. In the continuum limit this model was first studied in \cite{GZ} where it was written as a free fermion field theory 
with a particular boundary interaction term. As the boundary term is also quadratic in the bulk and the boundary fermion fields many quantities can be calculated exactly.  
Various aspects of this boundary field theory were subsequently studied in \cite{ CZ,Chat,Chat2,LMSK,KLM,MMM,me1, lattice_inhom1, lattice_inhom2,Toth,Toth2,Essler,Mir, new_g, me, Gaiotto}. 

The critical Ising model with a boundary magnetic field can be described as a perturbed boundary 
conformal field theory (BCFT). The critical Ising model is described as the Virasoro minimal model ${\cal M}(3,4)$ and it has 
three conformal boundary conditions corresponding to the free or fixed boundary spins in the underlying lattice model. 
The free boundary condition has a relevant boundary operator $\sigma_{B}$ of dimension 1/2 that describes the boundary spin. 
Let $S_{\rm BCFT}$ be the Euclidean action that describes the critical Ising model with free spin boundary condition 
on the upper half plane and let $z=x+iy$, $\bar z=x-iy$  be the complex coordinates. Switching on the boundary magnetic 
field amounts to deforming the action as 
\be \label{bmf}
S_{h} = S_{\rm BCFT} + h\!\int\limits_{-\infty}^{\infty}\!\! \sigma_{B}(x)dx \, . 
\ee
Here $h$ stands for the value of the boundary magnetic field and is a relevant coupling. The RG flow interpolates between the free 
and fixed spin boundary conditions.

The model (\ref{bmf}) can be generalised to a perturbation of a superposition of $N>1$ free boundary conditions. Let $i=1, \dots , N$ 
be the index labelling a copy of the free boundary condition. The boundary primaries on the superposition are dimension 1/2 fields $\sigma_{B}^{[ij]}$ and dimension zero fields $\chi^{[ij]}$ with $i,j=1, \dots , N$. 
Our conventions are such that the field insertion $\sigma_{B}^{[ij]}(x)$ has the $i$-th copy of the boundary condition to the right of $x$ and the $j$-th copy to the left, and similarly for $\chi^{[ij]}$.  We can then consider the most general relevant deformation 
\be \label{bmf_N} 
S_{h_{ij},m_{ij}} = S_{\rm BCFT} + \sum_{i,j=1}^{N} [h_{ij}\!\!\int\limits_{-\infty}^{\infty}\!\! \sigma_{B}^{[ij]}(x)dx + m_{ij}\!\!\int\limits_{-\infty}^{\infty}\!\! \chi^{[ij]}(x)dx] \, . 
\ee  
Here the coupling constants are arranged into  two $N\times N$ matrices: $h_{ij}$ and $m_{ij}$. 
For the perturbed theory to be unitary  these matrices have to be hermitian. 

The physical interpretation of the fields $\sigma_{B}^{[ij]}$  and  $\chi^{[ij]}$ may not be immediately clear. 
By taking the $N$ copies of the free boundary condition we have now endowed the boundary with 
additional degrees of freedom. As in the Kondo model we could try to interpret these degrees of freedom as those of an  
 impurity particle located at the boundary. Thus in the BCFT approach to Kondo model (see e.g. \cite{Affleck_lectures}) 
 the fields $\chi^{[ij]}$ describe the impurity spin degrees of freedom. As far as we can see this interpretation does not help with giving a physical meaning to 
 (\ref{bmf_N}). Instead we are offering a different kind of physical interpretation in terms of a non-constant boundary magnetic field with a fluctuating profile. As we will explain below the eigenvalues of $h_{ij}$ give possible values of the boundary magnetic field while  the values of the couplings $m_{ij}$ are related to statistical weights of different profiles.

Let us consider a deformed correlation function corresponding to  (\ref{bmf_N}) with some local insertions and some choice of 
vacua at $x\to \pm \infty$
\be \label{deformed_corr}
\langle \dots \rangle_{h,m;IJ} = \langle \exp(-S_{h_{ij},m_{ij}}) \dots \rangle_{0;IJ} \, .
\ee
Here the ellipsis stands for some local operator insertions and $\langle \enspace\rangle_{0;IJ}$ denotes 
the correlator in the unperturbed BCFT with the $I$-th copy of the boundary condition at $x\to \infty$ and the $J$-th copy 
at $x\to -\infty$. We can expand the exponential to obtain a series of boundary conformal perturbation theory. For 
boundary condition changing operators it was considered in \cite{Graham}. We choose the normalisations of the $\chi^{[ij]}$ fields so that they
satisfy the OPEs 
\bea \label{OPE1}
\chi^{[ij]}(x_{2}) \chi^{[kl]} (x_{1}) &= &\delta_{kj} \chi^{[il]}(x_1)   \, , \nonumber \\
\sigma_{B}^{[ij]}(x_{2}) \chi^{[kl]} (x_{1}) &= &\delta_{kj} \sigma_{B}^{[il]}(x_1)  \, , \nonumber \\
\chi^{[ij]}(x_{2}) \sigma_{B}^{[kl]} (x_{1}) &= &\delta_{kj} \sigma_{B}^{[il]}(x_1)   
\eea
where $x_{1}< x_{2}$. We also have a selection rule 
\be \label{OPE2}
\sigma_{B}^{[ij]}(x_{2}) \sigma_{B}^{[kl]} (x_{1}) = \delta_{jk} [\sigma_{B}(x_{2}) \sigma_{B}(x_{1}]^{[il]} \, , \quad x_{1}<x_{2} 
\ee
where the expression on the right hand side means that we should take the OPE in a single copy of the free spin boundary 
condition and then add the Chan-Paton indices $il$ to every term. The above OPEs considered inside the conformal perturbation 
expansion imply that the series can be rewritten as a path-ordered exponential 
\be \label{Pexp_form}
\langle \dots \rangle_{h,m;IJ} = \langle {\rm Pexp}(-S_{\rm BCFT} - [\int\limits_{-\infty}^{\infty}\!\! H(x)\sigma_{B}(x)dx + \int\limits_{-\infty}^{\infty}\!\! M(x)dx] )_{IJ} \dots \rangle_{0} \, .
\ee
where $H(x)$ and $M(x)$ stand for  the $h_{ij}$ and $m_{ij}$ matrices inserted at $x$ and the path ordering applies to these insertions. 
The correlator on the right hand side: $\langle \dots \rangle_{0}$ now refers to the VEV in a single copy of the free spin boundary condition.

The unperturbed BCFT possesses 
the ${\rm U}(N)$ Chan-Paton symmetry generated by the charges $\chi^{[ij]}$.  This symmetry is explicitly broken by the perturbation (\ref{bmf_N}) that results in emergence of some 
redundant boundary operators $\partial_{x}\chi^{[ij]}$ (see e.g.  appendix A in \cite{FK} for a more detailed discussion). 
Perturbations related by  unitary rotations of the coupling matrices are related by adding redundant operators to the action and are thus equivalent.
We can use this symmetry of the coupling space to diagonalise one of the coupling matrices.
Note that the operator $\sum_{i=1}\chi^{[ii]}$ 
commutes with all operators so that without loss of generality we can assume that $m_{ij}$ is traceless. 
One useful choice is to diagonalise $h_{ij}$ in which case $m_{ij}$ is generically some non-diagonal traceless hermitian matrix. 
Suppose the eigenvalues of $h_{ij}$ are the numbers $h_{1}, h_{2}, \dots , h_{N}$ and the index is the same as the index labelling the copies of the boundary condition.Then  it is not hard to see that the perturbed theory    
(\ref{bmf_N}) describes the critical Ising model on a half plane in the presence of a fluctuating piece-wise constant boundary  magnetic field. To get a precise picture consider the perturbed correlation function (\ref{deformed_corr}). Expanding in the powers of the $m_{ij}$ couplings we can consider terms with insertions 
\be
\prod_{i=1}^{m}  \chi^{[i_{m+1},i_{m}]}(x_{m})  \dots \chi^{[i_{3},i_{2}]}(x_{2}) \chi^{[i_{2},i_{1}]}(x_{1})
\ee
where $x_{1} < x_{2} <\dots < x_{m}$. The terms with boundary magnetic fields in the perturbation expansion are then selected according to OPEs  (\ref{OPE1}), 
(\ref{OPE2}) so that, after summing up the series in $h_i$, we obtain the boundary magnetic field with the value $h_{i_{k}}$ between $x_{k-1}$ and $x_{k}$ for $2\le k\le m$, and   with the values $h_{i_{1}}$ and $h_{i_{m+1}}$ for $x<x_{1}$ and $x>x_{m}$ respectively. 
A sample profile of this kind is depicted on Fig. \ref{profile}.

 \begin{center}
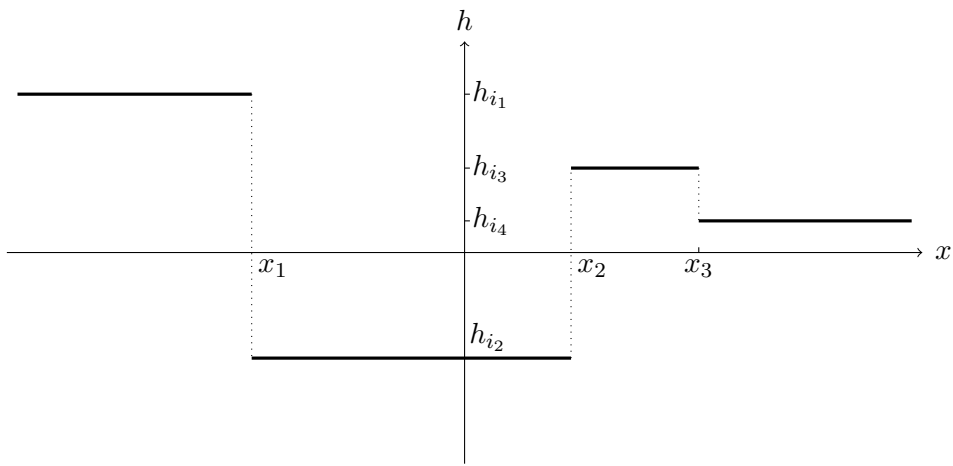
\begin{figure}[H]
\centering
\begin{tikzpicture}[scale=1.4]
\draw[->] (-4.3,0)--(4.3,0);
\draw[->] (0,-2)--(0,2);
\draw (4.5,0) node {$x$};
\draw (0,2.2) node {$h$};
\draw[very thick] (-4.2,1.5)--(-2,1.5);
\draw (0,1.5) --(0.05,1.5);
\draw (0.25,1.5) node {$h_{i_{1}}$};
\draw[dotted] (-2,1.5)--(-2,-1);
\draw (-1.8,-0.15) node {$x_{1}$};
\draw[very thick] (-2,-1)--(1,-1);
\draw (0.22,-0.8) node {$h_{i_{2}}$};
\draw[dotted] (1,-1)--(1,0.8);
\draw (1.2,-0.15) node {$x_{2}$};
\draw[very thick] (1,0.8)--(2.2,0.8);
\draw (0,0.8) --(0.05,0.8);
\draw (0.25,0.8) node {$h_{i_{3}}$};
\draw[dotted] (2.2,0.8)--(2.2,0.3);
\draw (2.2,0)--(2.2,0.05);
\draw (2.2,-0.15) node {$x_{3}$};
\draw[very thick] (2.2,0.3) --(4.2,0.3);
\draw (0,0.3) --(0.05,0.3);
\draw (0.25,0.3) node {$h_{i_{4}}$};

\end{tikzpicture}
\caption{Sample profile of the boundary magnetic field}
\label{profile}
\end{figure}
\end{center}
The summation over the indices $i_{1}, \dots, i_{m+1}$ and integration over the positions of the discontinuities $x_{1}, \dots, x_{m}$ 
describes an annealed disorder ensemble in which each coupling $m_{ij}$ with $i\ne $ plays the role of a fugacity variable for the discontinuity of 
the boundary magnetic field allowed to jump from the value $h_{j}$ to the value $h_{i}$. 
The couplings $m_{ii}$, $i=1,\dots , N$ correspond to shifts of the vacuum energy density in the corresponding sectors of the undeformed theory. They are additional simple deformations of the constant boundary magnetic field model. For $N=1$ such a perturbation drops out 
as it is proportional to the identity field but for $N>1$ and $m_{i} \ne m_{j}$, $i\ne j$ they lead to some non-trivial effects.  
We note that the quenched disorder case of the boundary 
magnetic field model was previously considered in \cite{Igloi}, \cite{Cardy_rand_bmf}, \cite{MMM}.  

To fully specify the model on a half plane we need to fix the boundary conditions at infinity that can be done for instance as in (\ref{deformed_corr}), (\ref{Pexp_form}) i.e. by fixing the boundary components in each limit. This does not treat all of the values  $h_{i}$ on equal footing. Alternatively we can consider averaging over the  different values of $I$ and $J$ in (\ref{deformed_corr}), (\ref{Pexp_form}). One natural way to do it is by putting the classical 
theory on a cylinder that is equivalent to considering the quantum theory at finite temperature $T$. 
Allowing all values $h_{i}$ to appear on a circle corresponds to taking a trace over the vacua labels $I$, $J$ so that the thermal partition function has the form 
\be \label{thermal} 
 Z_{\rm annealed} = C_{N} {\rm Tr} \langle {\rm Pexp}(-S_{\rm BCFT} - \int\limits_{0}^{\beta}\!\! H(\tau)\sigma_{B}(\tau)d\tau - \int\limits_{0}^{\beta}\!\! M(\tau)d\tau )  \rangle_{0}  
 \ee
 which is similar to a Wilson line. 
 Here $\tau$ is the Euclidean time coordinate that parameterises a circle of circumference $\beta=1/T$, 
 $C_{N}$ is a normalisation constant. If we choose $C_{N} = 1/N$ the partition function ({\ref{thermal}) describes 
 a single copy of the free boundary condition perturbed by a {\it non-constant} boundary magnetic field whose profile fluctuates along the boundary according to Fig.  \ref{profile} (adopted to a circle). The same model can be still described 
 as a  superposition of $N$  copies of the boundary boundary  perturbed by a collection of boundary fields with {\it constant} couplings 
 as in (\ref{bmf_N}). The $g$-theorem for boundary RG flows  \cite{gThm1}, \cite{gThm2} was proven for boundary theories  with constant couplings and thus applies to the latter kind of description.   In that description the $g$-factors are given by the boundary partition functions which are canonically normalised by the Cardy condition  \cite{Cardy}  applied to the irreducible components so that the normalisation constant $C_{N}=1$. In sections \ref{pf_sec}, \ref{RG_sec} when discussing the boundary RG flows we will choose this normalisation.

As shown in \cite{GZ} the model described by (\ref{bmf}) is integrable. Moreover one can deform the bulk theory by the energy density operator which corresponds to shifting the temperature away from critical value in the classical lattice model. The corresponding two-coupling model is still integrable. In the present paper we focus on the model described by  (\ref{bmf_N}) with  $N=2$. 
In addition to the critical case we will also consider (\ref{bmf_N}) with an additional bulk deformation by the energy density. In section 
\ref{free_ferm} we will show that the model can be described by a Lagrangian  theory containing the usual Majorana field and 3 real boundary fermions. Generically this Lagrangian contains a four-Fermi interaction term but  when $h_{1} = \pm h_{2}$ it is Gaussian. The case 
$h_{1} = -h_{2}$ is non-trivial and  to the best of our knowledge has not been studied before  (see however \cite{lattice_inhom1}, \cite{lattice_inhom2}    for a closely related work on the lattice). In section  \ref{R_sec} we 
derive and study the exact boundary reflection amplitudes describing the corresponding boundary integrable model that contains 4 parameters. In section \ref{pf_sec} we calculate the disc partition function for the same model in the bulk critical case. In section 
\ref{RG_sec} we use the disc partition function to study the details of the 3-coupling family of the RG boundary flows described by the $h_{1}=-h_{2}$ model critical in the bulk. In particular we elucidate the IR asymptotics of these flows. 
In section \ref{effective_sec} we work out the all orders infrared effective action. We finish the paper with some 
discussion in section \ref{conclude_sec}. The appendix contains details of calculations demonstrating the absence of conserved currents of spins 2 and 4 in the generic $N=2$ classical model.

\section{Free fermion description} \label{free_ferm}
 \setcounter{equation}{0}

 Let us consider the Ising field theory with zero magnetic field on a complex plane with coordinates $z=x+iy$, $\bar z= x-iy$. 
In the free fermion description the bulk action is\footnote{We use the conventions of \cite{yellow_book} in which 
$\epsilon=i:\!\!\psi\bar \psi\!\!:$, $C_{\sigma \sigma \epsilon} = 1/2$ and $m>0$ corresponds to the low temperature ordered phase.  } 
\be
S_{\rm bulk} = \frac{1}{2\pi}\! \iint \! [ \psi \bar \partial \psi +  \bar \psi \partial \bar \psi + im \psi \bar \psi ]dxdy
\ee
where 
\be
\partial \equiv \frac{\partial}{\partial z} = \frac{1}{2}\left( \frac{\partial}{\partial x} - i\frac{\partial}{\partial y}\right) \, , 
\quad \bar \partial \equiv \frac{\partial}{\partial \bar z} = \frac{1}{2}\left( \frac{\partial}{\partial x} + i\frac{\partial}{\partial y}\right) \, , 
\ee
and the coupling $m$ is proportional to $T_{c}-T$. We will assume that $m>0$ and the system is in the low temperature phase 
with two degenerate ground states. 

Suppose now that we have a simply connected region ${\cal B}$ on the complex plane with boundary ${\cal D}=\partial {\cal B}$  that is connected and is  described as a curve $z=Z(t)$, $\bar z = \bar Z(t)$. Following \cite{CZ} we assume that the parameterisation is chosen in  such a way that 
\be
e(t)\bar e(t) = 1 \, , \quad \mbox{where  } e(t) = \frac{dZ}{dt}(t)\, , \enspace \bar e(t) = \frac{d\bar Z}{dt}(t)
\ee
and the unit normal vector $(ie,-i\bar e)$ points inside ${\cal B}$. With these conventions a suitably normalised boundary spin operator 
can be introduced as \cite{CZ} 
\be \label{sigmaB}
\sigma_{B}(t) = i a(t) (e^{1/2}(t) \psi(t) + \bar e^{1/2} \bar \psi(t)) 
\ee
 where $\psi(t)\equiv \psi(Z(t))$, $\bar \psi(t)\equiv \bar \psi(\bar Z(t))$ and $a(t)$ is the boundary fermion that is a real fermionic boundary field with the two-point function 
 \be
 \langle a(t) a(t')\rangle = \frac{1}{2} {\rm sign}(t-t') \, . 
 \ee
The Ising field theory on ${\cal B}$ with a boundary magnetic field on ${\cal D}$ is then described by the action 
\be \label{Sfree}
S= \frac{1}{2\pi}\! \iint\limits_{\cal B} \! [ \psi \bar \partial \psi +  \bar \psi \partial \bar \psi + im \psi \bar \psi ]dxdy + 
\int [\frac{1}{2} a\dot a - \frac{i}{4\pi}\psi \bar \psi(t)] dt + h\!\int \!\sigma_{B}(t) dt 
\ee
where $\dot a = \frac{da}{dt}$. For the upper half plane we can choose $Z(t) = t$, $\bar Z(t) =t$ and $e(t)=\bar e(t)=1$.  The action  (\ref{Sfree}) then
gives the standard bulk equations of motion and the boundary conditions 
\bea
&& \psi(t)-\bar \psi(t) = 4\pi h a(t) \, , \nonumber \\
&& \dot a = -ih(\psi(t) + \bar \psi(t)) \, .
\eea
 Combining the two equations we can obtain a closed boundary condition on the $\psi$ and $\bar \psi$ fields: 
 \be
 \partial_{t} \psi(t) + i \lambda \psi(t) = \partial_{t} \bar \psi(t) -i\lambda \bar \psi(t)
 \ee
where 
\be
\lambda = 4\pi h^2 \, .
\ee
As first noticed in \cite{CZ} this boundary condition means that the field $\chi(z) \equiv (\partial + i\lambda)\psi(z)$ 
can be extended to a holomorphic field on the entire complex plane.  The fact that $\chi(z)$ depends only on the square of the coupling means that the gluing condition remains the same in the presence of discontinuities in the boundary magnetic field at which it changes the sign. It is not hard to show that in this case the field $\chi(z)$ is meromorphic on the plane with simple poles at the discontinuities. 
This singles out this type of discontinuity and suggests integrability of the theory with such a fixed profile. On the lattice such models have been worked out in terms of free fermions in \cite{lattice_inhom1}, \cite{lattice_inhom2}. In this paper we are mainly interested in the case when the discontinuities are dynamic in the sense of an annealed ensemble in which the discontinuities are governed by a fugacity coupling. We proceed next to a free fermion description of perturbations of the form (\ref{bmf_N}) defined on superpositions.  


The use of boundary fermions to describe Chan-Paton interactions and Wilson loops in string theory is a well known trick 
 \cite{Marcus1} , \cite{Marcus2}. 
From the point of view of the conformal field theory both the bulk fermionic fields $\psi(z)$ and $\bar \psi(\bar z)$ and the boundary fermion $a(t)$ can be described in terms of topological defects \cite{duality}, \cite{Runkel_Watts}.  The critical Ising model has three topological defects labeled by the primary fields. They can end on the conformal boundary conditions. The $\epsilon$-defect is the spin reversal symmetry defect. The bulk free fermion fields $\psi$ and $\bar \psi$ are located at the tip of the $\epsilon$ defect. This defect can also  end on the free boundary condition forming either a topological junction that (with an appropriate choice of normalisation) corresponds to the boundary fermion field $a$ or via a non-topological junction given by a weight $1/2$ fermionic field that (up to normalisation) is given by the $\psi$ (or $\bar \psi$) field inserted on the boundary\footnote{Recall that any holomorphic or anti-holomorphic field cannot have singularities at the boundary, with a conformal boundary condition, and thus can be put on the boundary. That gives a boundary field of the same conformal weight.}. Given two such junctions: one topological and one non-topological, we can merge them together. In the bulk this results in fusing the $\epsilon$-defect with itself that gives the identity defect and on the boundary that gives a (bosonic) ``ordinary'' boundary field of dimension 1/2. This boundary field is proportional to the boundary spin field $\sigma_{B}$. This process depicted on Figure \ref{junction_pic} explains formula (\ref{sigmaB}). Such a relation along with other bulk-to-boundary and boundary OPEs involving fermionic fields has been worked out in \cite{Runkel_Watts} from first principles.

\begin{figure}[H]
\begin{center}
\begin{tikzpicture}[>=latex,scale=2]
\draw [white,pattern=my north east lines,  line space=5pt, pattern color=black] (-1,0) rectangle (1,-0.1) ;
\draw[very thick, red] (-0.2,0)--(-0.2,1);
\draw (-0.3,0.5) node {$\epsilon$};
\draw[very thick, red] (0.2,0)--(0.2,1);
\draw (0.1,0.5) node {$\epsilon$};
\draw[blue] (-0.2,0) node {$\bullet$} ;
\draw (-0.2,-0.2) node {$a$};
\draw (0.2,-0.2) node {$\psi$};
\draw[green] (0.2,0) node {$\bullet$} ;
\draw[very thick] (-1,0)--(1,0);
\draw (1.2,0) node {$=$}; 
\draw [white,pattern=my north east lines,  line space=5pt, pattern color=black] (1.4,0) rectangle (3,-0.1) ;
\draw[very thick] (1.4,0)--(3,0);
\draw (2.2,0) node {$\bullet$};
\draw (2.2,-0.2) node {$\alpha \sigma_{B}$};
\end{tikzpicture}
\end{center}
\caption{Merger of two defect junctions producing a boundary magnetic field operator. $\alpha\ne 0$ is a normalisation related constant. }
\label{junction_pic}
\end{figure}
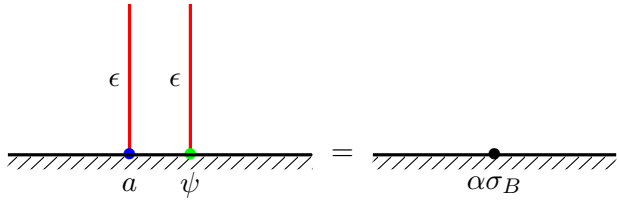

Let us consider now a superposition of $N>1$ copies of the free spin boundary conditions. The $\epsilon$-defect has a topological and a non-topological junction between any pair of copies. This gives rise to the junction fields $a^{[ij]}$ and $\psi^{[ij]}$ which have dimension 0 and dimension 1/2 respectively. Like for boundary fields our notation is such that $a^{[ij]}$ when inserted at a boundary point of the upper half plane has the $j$-th copy of the boundary condition  to the left and the $i$-th copy to the right, and similarly for $\psi^{[ij]}$.
Choosing an appropriate normalisation of the $\psi^{[ij]}$ fields we now have 
\be \label{psi_lim}
\lim_{z \to \tau \in {\mathbb R}} \psi(z) = \sum_{i=1}^{N} \psi^{[ii]}(\tau) 
\ee
so that only the combination on the right hand side of this equation can be moved from the boundary into the bulk. 
We choose the normalisations of the $a^{[ij]}$ fields so that merging of the topological junctions between themselves is given by the following (product) OPE rule 
\be \label{aa_rel}
 a^{[ij]} a^{[kl]} = \delta_{jk} \chi^{[il]} \, .
\ee
Furthermore we choose the normalisations of the $\sigma_{B}^{[ij]}$ fields in such a way that 
\be
\sigma_{B}^{[ij]}(t) = i a^{[ik]}(t) (e^{1/2}(t) \psi^{[kj]}(t) + \bar e^{1/2} \bar \psi^{[kj]}(t)) \quad \mbox{for any } k \, .
\ee
This allows us to write 
\be
\sigma_{B}^{[ij]}(t) = i a^{[ij]}(t) (e^{1/2}(t) \psi(t) + \bar e^{1/2} \bar \psi(t)) = -i (e^{1/2}(t) \psi(t) + \bar e^{1/2} \bar \psi(t)) a^{[ij]}(t)
\ee
where the $\psi(t)$ and $\bar \psi(t)$ fields on the right hand side are given by (\ref{psi_lim}). To obtain a functional integral representation for 
the theory described by   (\ref{bmf_N}) we need a set of anti-commuting boundary fields. Although one can choose such a set for any value of $N$ from now on we will specialise to the $N=2$ case on which the present paper is focused. Consider the combinations 
\bea
a &=& a^{[11]}-a^{[22]} \, , \nonumber \\
b&=& i(a^{[21]}-a^{[12]}) \, , \nonumber \\
c&=& a^{[21]} + a^{[12]} \, .
\eea 
Using (\ref{aa_rel}) we check that these fields satisfy 
\be \label{abc_rels}
\{a,b\} = \{a,c\}=\{b,c\} = 0 \, , \qquad a^2 = b^2 = c^2 = \chi^{[11]} + \chi^{[22]} \equiv {\bf 1} \, . 
\ee
We also find the following relations
\bea
& iab = \chi^{[12]} + \chi^{[21]}\, , \quad ibc = \chi^{[11]} - \chi^{[22]} \, , \quad ac = \chi^{[12]}-\chi^{[21]} \, ,  &\nonumber \\
& abc = -i(a^{[11]} + a^{[22]}) &
\eea
and 
\be
 a^{[11]} = \frac{1}{2}(a + iabc) \, , \quad a^{[22]} = \frac{1}{2}(-a + iabc) \, .
\ee
We note that the fields $a,b,c$  are hermitian that follows from the conjugation relations 
\be
(\chi^{[ij]})^{\dagger} = \chi^{[ji]} \, . 
\ee 
The above relations mean that in the gauge in which  $h_{ij}={\rm diag}(h_{1}, h_{2})$ we can represent the $N=2$ 
theory   (\ref{bmf_N}) as a functional integral 
\be
\int e^{-S} D[\psi]D[ \bar \psi] D[a] D[b] D[c]
\ee
with the action functional 
\bea \label{S2}
&& S=\frac{1}{2\pi}\! \iint\limits_{\cal B} \!    (\psi\bar \partial \psi + \bar \psi \partial \bar \psi + im \psi \bar \psi)dxdy + \int\limits_{-\infty}^{\infty}\!\!  \Bigl[ -\frac{i}{4\pi}\psi\bar \psi 
+ \frac{1}{2}a\dot a + \frac{1}{2}b\dot b + \frac{1}{2}c\dot c \nonumber \\
&& +  (ih_{-}a + h_{+}abc)(e^{1/2} \psi + \bar e^{1/2} \bar \psi)  + m_{1}iab + m_{2}ibc \Bigr]  dt
\eea
Here 
\be \label{hh}
h_{+} = -\frac{h_{1} + h_{2}}{2} \, , \quad h_{-} = \frac{h_{1}-h_{2}}{2} \, . 
\ee
and\footnote{Here  without loss of generality we assume that $m_{1}$ is real as that can be achieved by a residual unitary transformation that preserves the diagonal matrix $h_{ij}$.}
\be \label{mm} 
(m_{ij}) = \left(\begin{array}{rr} m_{2}& m_{1} \\
m_{1}& -m_{2} \end{array} \right)  \, . 
\ee
We immediately observe that when $h_{+}=0$ the theory (\ref{S2}) is Gaussian. 
From now on we will assume that $h_{+}=0$ postponing the discussion of the general $h_{-}h_{+} \ne 0$ case until section  \ref{conclude_sec}. When $h_{+}=0$, 
considering the model on the upper half plane we obtain the following bulk equations of motion and boundary conditions
\be\label{bulk_eqmo}
\bar \partial \psi + \frac{im}{2}\bar \psi = 0 \, , \qquad \partial \bar \psi - \frac{im}{2}\psi = 0 \, , 
\ee
\bea \label{bcs}
 \psi(t) - \bar \psi(t) &=& 4\pi h_{-} a(t) \, ,  \nonumber \\ 
 \dot a(t) &=& -ih_{-}(\psi(t) + \bar \psi(t)) - m_{1}ib(t)   \, ,  \nonumber \\  
 \dot b(t) &=& m_{1}ia(t) - m_{2}ic(t)  \, ,  \nonumber \\ 
 \dot c(t) &=& m_{2}ib(t) \, .  
\eea
By taking more derivatives of the equations in (\ref{bcs}) we can obtain a closed boundary condition 
on the $\psi$ and $\bar \psi$ fields that looks as follows 
\be \label{bcpsi}
\partial_{t}^{3}\psi + i\lambda\partial_{t}^2\psi - (m_{1}^2+m_{2}^2)\partial_{t} \psi -im_{2}^2\lambda \psi 
= \partial_{t}^{3}\bar \psi - i\lambda\partial_{t}^2\bar \psi - (m_{1}^2+m_{2}^2)\partial_{t} \bar \psi +im_{2}^2\lambda \bar \psi 
\ee
  where 
  \be
  \lambda \equiv \lambda_{-} = 4\pi h_{-}^2  \, .
  \ee

\section{Exact boundary reflection amplitudes }  \label{R_sec}
 \setcounter{equation}{0}
 
To find the boundary reflection amplitude corresponding to (\ref{S2}) we  put the model onto a left half-plane: ${\rm Re} z \le 0$, and 
parameterise the boundary by $t= y={\rm Im} z$. This gives $e=\omega$, $\bar e = \bar \omega$ where 
\be
\omega = e^{i\frac{\pi}{4}}
\ee
and the boundary conditions are   given by equations (\ref{bcs}), (\ref{bcpsi}) in which $\psi$ and $\bar \psi$ are 
replaced by $\omega \psi$ and $\bar \omega \bar \psi$ respectively. 
The canonical quantisation expansion of the fields $\psi$ and $\bar \psi$ corresponding to (\ref{bulk_eqmo}) has the form 
\bea \label{mode_exp}
\psi(x,y) = \int\limits_{-\infty}^{\infty}\!\! d\theta \Bigl[  \omega e^{\theta/2} A(\theta) e^{-my\cosh\theta + imx\sinh\theta} + \bar \omega e^{\theta/2} A^{\dagger}(\theta) e^{my\cosh\theta - imx\sinh\theta} \Bigr] \, ,  &&\nonumber \\
 \bar \psi(x,y) =  \int\limits_{-\infty}^{\infty}\!\! d\theta \Bigl[  \omega e^{-\theta/2} A(\theta) e^{-my\cosh\theta + imx\sinh\theta} - \bar \omega e^{-\theta/2} A^{\dagger}(\theta) e^{my\cosh\theta - imx\sinh\theta} \Bigr]  &&
\eea 
where $y$ is the Euclidean time, $\theta$ is the rapidity variable and $A(\theta)$, $A^{\dagger}(\theta)$ are the creation and annihilation operators satisfying 
\be
\{ A(\theta) , A^{\dagger}(\theta')\} = \delta(\theta - \theta') \, .
\ee 
Following \cite{GZ} the boundary can be treated as an infinitely heavy particle located at $x=0$. The vacuum state for the $t$-quantisation is formally given by the action of a boundary creation operator $B$ on the bulk theory vacuum:
\be
|0\rangle_{B} = B|0\rangle \, .
\ee 
The reflection coefficient $R(\theta)$ is then defined by the relation
\be
A^{\dagger}(\theta)B = R(\theta) A^{\dagger}(-\theta)B \, . 
\ee

Substituting the expansions (\ref{mode_exp}) into the boundary condition (\ref{bcpsi}) we obtain the reflection matrix 
\be \label{R1}
R(\theta) = i\frac{p^{0}((p^{0})^2 - m_{1}^2-m_{2}^2)\cosh\left(\frac{\theta}{2}-i\frac{\pi}{4}\right) -i\lambda((p^{0})^2-m_{2}^2)\sinh\left(\frac{\theta}{2}-i\frac{\pi}{4}\right)}{p^{0}((p^{0})^2 - m_{1}^2-m_{2}^2)\sinh\left(\frac{\theta}{2}-i\frac{\pi}{4}\right) +i\lambda((p^{0})^2-m_{2}^2)\cosh\left(\frac{\theta}{2}-i\frac{\pi}{4}\right)}
\ee
where
\be
p^{0}=m\cosh(\theta) \, .
\ee
Another useful form for this amplitude is 
\be \label{R2}
R(\theta) = i\tanh\left(\frac{\theta}{2}-i\frac{\pi}{4}\right) \frac{(\cosh^2(\theta)-a_1 - a_2)(\sinh(\theta)+i) 
-ib(\cosh^2(\theta)-a_2)}{(\cosh^2(\theta)-a_1 - a_2)(\sinh(\theta)-i) 
+ib(\cosh^2(\theta)-a_2)}
\ee
where we introduced dimensionless parameters 
\be
a_1=\frac{m_1^2}{m^2} \, , \quad a_2=\frac{m_2^2}{m^2}\, , \quad b=\frac{\lambda}{m} \, .
\ee
It is straightforward to check that the reflection amplitude given in  (\ref{R1}), (\ref{R2}) satisfies the unitarity and boundary crossing symmetry: 
\be
R(\theta) R(-\theta)=1 \, , \qquad R\left( i\frac{\pi}{2} - \theta\right) = -R\left( i\frac{\pi}{2} + \theta\right) \, .
\ee
Of particular interest are poles of $R(\theta)$ located in the physical region: $\theta = i \xi$, $\xi \in [0,\pi]$. They correspond to boundary bound states. To find them it is convenient to set $\sinh(\theta) = ix$ then the denominator in (\ref{R2}) up to a factor of $-i$ is given by
the polynomial 
\be
q(x) = x^3 + x^2(b-1) - x(1-a_1-a_2) + a_{2}b+1 - a_{1} - a_{2} - b  \, .
\ee
This polynomial vanishes when we have 
\be
(x-1+b)(x^2-1+a_2) = a_{1}(1-x) \, .
\ee
From this representation we see that, assuming $a_{1}>0$, $a_{2}>0$, $b>0$, we always have a single bound state, that is a pole in the physical interval $x\in [0,1]$, when $a_{1}>(1-a_2)(1-b)$. Also when $a_{1} <(1-a_2)(1-b)$ and $a_{2}< 1$ we have two distinct bound states and when  $a_{1} <(1-a_2)(1-b)$ and $a_{2}>1$ we have no bound states. When  $a_{1}=(1-a_{2})(1-b)>0$ there is no bound state unless $a_{2}<1$ and $b<1$ in which case there is a single bound state (note that the zero of the denominator of $R$ at $x=0$ is cancelled by the zero of the numerator). As argued in \cite{GZ} (see also discussions in \cite{KLM} and \cite{Essler}) the boundary bound states which are present for sufficiently small boundary couplings correspond to the deformed degenerate ground states.  In the undeformed model we are considering the vacuum state on a half-plane is four times degenerate. It is interesting to note then that we can only have at most  two boundary bound states rather than three expected naively (plus the vacuum). It would be interesting to obtain an intuitive  physical picture explaining this. 

The remaining special cases when one of the three couplings vanishes can be described as follows. 
When $b=0$ the reflection coefficient (\ref{R2}) reduces to the one for the free boundary condition:
\be
R_{\rm free}(\theta) = i \coth\left( \frac{\theta}{2}-i\frac{\pi}{4} \right) \, .
\ee 
This is because in this case 
we only have the boundary identity fields switched on which decouple from the bulk fermion field.
When $a_{1}=0$ the model factorises into two decoupled boundary conditions with the boundary magnetic field given by $h_{-}$ and $-h_{-}$ and additional shifts of vacuum energy given by $m_{2}$ and $-m_{2}$. The vacuum energy shifts decouple from the bulk fermion and we have the reflection amplitude of the usual boundary magnetic field model \cite{GZ}
\be
R_{h}(\theta) = -i\tanh\left(\frac{\theta}{2}-i\frac{\pi}{4}\right) \frac{1-b-i\sinh(\theta)}{1-b+ i\sinh(\theta)} \, . 
\ee
When $a_{2}=0$ we obtain a two coupling model with reflection amplitude that can be written as 
\be \label{Rab}
R_{a,b}(\theta) = i\coth\left( \frac{\theta}{2}-i\frac{\pi}{4} \right) \frac{\sinh^2(\theta) -ib\sinh(\theta) + 1-a_1-b}{\sinh^2(\theta) +ib\sinh(\theta) + 1-a_1-b} \, .
\ee
 For $a_1\ne 0$ this amplitude has a pole at $\theta = i\frac{\pi}{2}$. This pole is absent when all three couplings are switched on (we actually get a zero at $\theta = i\frac{\pi}{2}$ in that case).  As discussed in \cite{GZ} such a pole, which is present for the free spin boundary condition,  means that the boundary state has a contribution proportional to the momentum zero one-particle state. For the free spin boundary condition this is interpreted as the spin reversal symmetry defect attached to the boundary. 
 In our case we can interpret this one-particle state as a superposition   of two spin-reversal symmetry defects  attached to two copies of the free boundary condition as described by the junction field
 \be
c=a^{[12]} + a^{[21]} \, . 
 \ee
 As can be seen from (\ref{abc_rels})  this junction field commutes with the perturbation when $m_{2}=0$. 
 
In addition to the above pole, the reflection amplitude $R_{a,b}$ has another pole\footnote{As usual there is also the  crossing symmetric pole at $\theta=i(\pi - \xi_{*})$.}  at $\theta =i\xi_{*}$ 
with
\be
\xi_{*} = \arcsin\left(  -\frac{b}{2} + \frac{1}{2}\sqrt{b^2 + 4(1-b-a_{1})} \right)
\ee
which is located on the physical interval $\xi\in[0,1]$ provided $a_1+b<1$.  
  We plan to discuss correlation functions and the physical interpretation of the boundary bound states we found in this section in a separate publication.

\section{Exact disc partition functions in the critical case} \label{pf_sec}
 \setcounter{equation}{0}
 
In this section we calculate the disc partition function for the three-coupling  model (\ref{S2}) in the case of conformal bulk that is when $m=0$. We aim  to find the boundary entropy and investigate the boundary RG flows. 

We start by putting the model (\ref{S2}) on a cylinder with complex coordinate $z=x+i\tau$ where $x\ge 0$ is the coordinate along the cylinder, $ \tau \in [0,\beta]$ is the Euclidean time, $\beta$ is the inverse temperature. To calculate the partition function on a disc we assume that on the cylinder the boundary condition at $x\to \infty$ corresponds to the bulk theory conformal vacuum and thus we take all fermionic fields: $\psi$, $\bar \psi$, $a,b,c$ in the NS sector that is antiperiodic in $\tau$. 
At the beginning we will treat the functional integral formally introducing a regularization and renormalisation later. 
The integral over $\psi$ and $\bar \psi$ with fixed boundary values is Gaussian and hence localises on the solutions to the bulk equations of motion decaying in the bulk.  We can write these solutions as the following mode expansions
\be
\psi(z) = \sum\limits_{k\in 1/2+{\mathbb Z}; k<0} \psi_{k} e^{\frac{2\pi}{\beta}k z} \, , \quad 
\bar \psi(\bar z) =  \sum\limits_{k\in 1/2+{\mathbb Z}; k>0}  \bar \psi_{k} e^{-\frac{2\pi}{\beta} k\bar z} \, .
\ee
When restricted to the boundary $x=0$ these solutions give anti-periodic functions
\be
\psi_{b} (\tau) \equiv \psi(i\tau) \, , \qquad \bar \psi_{b} (\tau) \equiv \bar \psi(-i\tau)\, . 
\ee
We further use the Fourier expansions 
\be
a(\tau)=\sum\limits_{k\in 1/2 + {\mathbb Z}} a_{k} e^{i\frac{2\pi}{\beta} k \tau}  \, , \qquad b(\tau)=\sum\limits_{k\in 1/2 + {\mathbb Z}} b_{k} e^{i\frac{2\pi}{\beta} k \tau} \, , 
\ee
\be
c(\tau)=\sum\limits_{k\in 1/2 + {\mathbb Z}} c_{k} e^{i\frac{2\pi}{\beta} k \tau}  \, .
\ee
Integrating over the fermions  $\psi$ and $\bar \psi$ with the fixed values on the boundary
we obtain up to the  divergent determinant factor the following fermionic path integral 
\bea
&& Z_{\rm disc}  = \int D[\psi_{b}] D[\bar \psi_{b}] D[a]D[b]D[c]  e^{-S_{a,b,c}} \nonumber \\
&& \times \exp\left( -i\frac{\beta}{4\pi} \sum_{k<0} \psi_{k}\bar \psi_{-k} 
 + ih\beta \sum_{k<0} (\omega \psi_{k} a_{-k} + \bar \omega \bar \psi_{-k} a_{k}) \right)
\eea
where
\be
S_{a,b,c} = \int d\tau  \Bigl[ \frac{1}{2}a\dot a + \frac{1}{2}b\dot b + \frac{1}{2}c\dot c +   + m_{1}iab + m_{2}ibc \Bigr]  \, . 
\ee
Removing the determinant factor here amounts to subtracting the bulk CFT vacuum energy density so that what remains corresponds to the disc partition function.  Integrating out the boundary functions $\psi_{b}(\tau)$ and $\bar \psi_{b}(\tau)$ gives 
\be
 Z_{\rm disc}  = D_{1} \int D[a]D[b]D[c]  e^{-S_{a,b,c}} \exp\left(  i\sum_{k<0} a_{-k}a_{k} (\lambda \beta -2\pi k)  \right)
\ee
where 
\be 
D_{1} = \prod_{n=1}^{\infty} \left( -i \frac{\beta}{4\pi}\right) 
\ee
is a formal expression for the functional determinant. As this factor does not depend on any coupling constants we remove it from the renormalised partition function. What remains now is a path integral over the boundary fermions: $a,b,c$. 
We normalise this path integral in such a way that in the Hamiltonian quantisation it amounts to taking the   
trace over the degenerate vacuum space. The remaining Gaussian integrations now produce the following formal expression 
\be \label{Zdisc2}
 Z_{\rm disc}  = C \prod\limits_{m=0}^{\infty} \Bigl[ \left( 1 + \frac{\alpha}{m+1/2}\right)\left( 1 + \frac{\nu_{2}}{(m+1/2)^2}\right) + 
 \frac{\nu_1}{(m+1/2)^2}   \Bigr]
\ee
where 
\be
\alpha = \frac{\lambda \beta}{2\pi} = 2\beta h_{-}^2 \, , \qquad \nu_{1} = \frac{\beta^2 m_{1}^2}{4\pi^2} \, , \quad 
\nu_{2} = \frac{\beta^2 m_{3}^2}{4\pi^2} 
\ee
and
the factor 
\be
C = \Bigl[\prod_{k=0}^{\infty} (-2\pi i) (k+1/2) \Bigr]^3 
\ee
comes from the kinetic terms of the boundary fermions. As this factor should give the trace of the identity in the vacuum space we set it equal to 2:
\be
C_{\rm ren} = 2 \, .
\ee
The remaining infinite product in (\ref{Zdisc2})  we regularise as follows
\bea
&& Z_{\rm disc}^{(\epsilon)} = 2 [ (1+2\alpha)(1+4\nu_{2}) + 4\nu_{1}]  \nonumber \\
&& \times \prod\limits_{m=1}^{\infty} \Bigl[ \left( 1 + \frac{\alpha}{m+1/2}\right)\left( 1 + \frac{\nu_{2}}{(m+1/2)^2}\right) + 
 \frac{\nu_1}{(m+1/2)^2}   \Bigr] e^{ \frac{\alpha}{m}(e^{-\epsilon m}-1)}  
\eea
where $\epsilon>0$ is a regulator. The regulated expression can be written in a manifestly finite form 
\be \label{Zreg}
Z_{\rm disc}^{(\epsilon)} = \frac{2\pi^{3/2}e^{-\alpha \ln(1-e^{-\epsilon}) }}{\Gamma(1/2-k_{1})\Gamma(1/2-k_2)\Gamma(1/2-k_3)} 
\ee
where $k_{1}$, $k_{2}$ and $k_{3}$ are roots of the polynomial 
\be \label{pz}
p(z) \equiv z^{3} + \alpha z^2 + (\nu_{1} + \nu_{2}) z + \alpha \nu_{2} = (z-k_{1})(z-k_{2})(z-k_{3}) 
\ee
and are thus radical functions of the couplings. The regularised expression (\ref{Zreg}) has a logarithmic divergence in 
the exponential in the limit $\epsilon \to 0$. In perturbation theory this divergence emerges at the second order when 
 we  integrate the two-point function $\langle \sigma_{B}(\tau) \sigma_{B}(\tau')\rangle$ of the boundary spin operator. 
 Subtracting this divergence we obtain a renormalised value of the partition function 
 \be \label{Z3ren}
 Z_{\rm disc, ren} = \frac{2\pi^{3/2}e^{\alpha \ln(\beta \mu) }}{\Gamma(1/2-k_{1})\Gamma(1/2-k_2)\Gamma(1/2-k_3)}
 \ee
where $\mu$ is the subtraction scale which has dimension of mass. When the couplings $\alpha$, $\nu_{1}$, $\nu_{2}$ 
are all zero we have $k_{1}=k_{2}=k_{3}=0$ and $ Z_{\rm disc, ren} = 2$ that is the correct $g$-factor of two copies of the free spin boundary condition. The subtraction scale $\mu$ is in principle arbitrary but can be fixed  by requiring that the partition function 
tends to a finite limit in the far infrared where $\beta \to \infty$. 
For example when $\nu_{2}=\nu_{1}=0$ this requirement fixes the subtraction point to be  
\be
\mu \beta = \frac{\alpha}{e} \, .
\ee 
In this case 
the model essentially reduces to the boundary magnetic field model. Up to a factor of 2 the 
partition function is the same as  the one computed in  \cite{Chat} 
\be
Z_{h} = \frac{\sqrt{2\pi}}{\Gamma(\alpha + 1/2)} \left( \frac{\alpha}{e}\right)^{\alpha} \, .
\ee
In the generic case the infrared asymptotic of  (\ref{Z3ren}) is 
\be
 Z_{\rm disc, ren} \sim {\rm Const} \exp\left[\alpha \ln(\beta \mu)  + \alpha + \sum_{i} k_{i}{\rm Log}\left( \frac{1}{2}-k_{i} \right) \right]
\ee
 and we can set 
\be \label{fixed_subtr}
\mu \beta = \exp\left( - 1 - \sum_{i} \frac{k_{i}}{\alpha} {\rm Log} (-k_{i}) \right) \, .
\ee
Here ${\rm Log}(z)$ stands for the principal branch of the complex natural logarithm function. The above expressions are real as the roots $k_{i}$ are either all real and negative or are comprised of a complex pair and one real negative root.  

In the following we will be interested in the boundary entropy rather than the partition function itself. In the boundary entropy the subtraction scale $\mu$ drops out so we do not need  to assume  it is fixed as in  (\ref{fixed_subtr}) or otherwise. 

\section{RG flows} \label{RG_sec}
 \setcounter{equation}{0}
 
The model (\ref{S2}) describes a relevant perturbation of the superposition of two copies of the free spin boundary condition $(f)$. 
Its $g$-factor \cite{gThm1} equals $2$. The fixed spin conformal boundary conditions $(+)$ and $(-)$ each have the $g$-factor  
$g_{+}=g_{-}=\frac{1}{\sqrt{2}}$. The $g$ theorem \cite{gThm1}, \cite{gThm2} implies that the boundary entropy monotonically decreases and thus the RG flow starting with $2(f)$ can end up in any of the   8 fixed points: $(f)\oplus(+)$, $(f)\oplus(-)$,  $(+)\oplus (-)$, $2(+)$, $2(-)$, $(f)$, $(+)$, $(-)$. 
We show below that the three parameter subspace of flows with $h_{+}=0$ ends on half of these possible fixed points: $(+)\oplus (-)$, $(f)$, $(+)$, $(-)$. 

The boundary entropy of (\ref{S2}) with $h_{+}=0$ is given by 
\be
s= \left( 1 - \beta \frac{\partial}{\partial \beta} \right) \ln\! Z_{\rm disc, ren} 
\ee
where $Z_{\rm disc, ren}$ is given in (\ref{Z3ren}). The large $\beta$ asymptotic of $s$ depends on the values of the couplings. 
Since $k_{1}k_{2}k_{3}=-\alpha \nu_{2}$, when $\alpha \nu_{2} \ne 0$ the roots $k_{i}$ are all non-vanishing. It also follows from (\ref{pz}) 
that 
\be
\beta \frac{\partial k_{i}}{\partial \beta} = k_{i}
\ee
and hence in the case at hand each $k_{i}$ tends to infinity homogeneously in $\beta$. 
Using the asymptotic expansion\footnote{This expansion can be obtained by combining formulae (1.18.12),  (1.13.14), (1.12.2) and (1.12.21) from \cite{BE}.} 
\be
\ln \Gamma(1/2 + z) = z\ln(z) - z + \frac{1}{2}\ln(2\pi) + \sum_{n=1}^{\infty} \frac{B_{2n}(2^{1-2n}-1)}{2n(2n-1)z^{2n-1} }
\ee
where $B_{m}$ denotes the Bernoulli numbers, we obtain the asymptotic expansion 
\be
s = -\frac{1}{2}\ln 2 + \sum_{i=1}^{3}\sum_{n=1}^{\infty} \frac{B_{2n}(1-2^{1-2n})}{(1-2n)k_{i}^{2n-1}} \, .
\ee
This gives the infrared fixed point $g$-factor to be 
\be g_{\rm IR} =\frac{1}{\sqrt{2}}
\ee  
that is the value of the $(+)$ or $(-)$ boundary condition. 
On physical grounds, which of the two fixed points the flow approaches depends on the sign of the product $h_{-}m_{2}$. (The coupling $m_{2}$ shifts the ground energy density  by   $-m_{2}$ for the portion of the boundary with the magnetic field $h_{-}$ and by the opposite value the portion perturbed by $-h_{-}$. This favours the value for which the shift is negative.)

We next consider the cases when $\alpha \nu_{2}=0$. 
 When $\nu_{2}=0$, $\alpha\ne 0$, $\nu_{1}\ne 0$ the polynomial (\ref{pz}) takes the form 
\be
p(z) = z(z^2 + \alpha z + \nu_{1})
\ee
and we can set 
\be
k_{1} = 0\, , \quad k_{2} = -\frac{\alpha}{2} + \frac{1}{2}\sqrt{\alpha^2 - 4\nu_{1}} \, , \quad 
 k_{3} = -\frac{\alpha}{2} - \frac{1}{2}\sqrt{\alpha^2 - 4\nu_{1}} \, .
\ee
The boundary entropy in this case has the small temperature asymptotic expansion 
\be
s = \sum_{i=1}^{2}\sum_{n=1}^{\infty} \frac{B_{2n}(1-2^{1-2n})}{(1-2n)k_{i}^{2n-1}}
\ee
that corresponds to 
\be
g_{\rm IR} = 1
\ee
and hence the boundary condition flows to the (single copy of) free spin boundary condition. 

When $\nu_{1}=\nu_{2}=0$, $\alpha\ne 0$ we have two factorised boundary magnetic field flows from free to fixed spin and the endpoint of the flow is 
the superposition $(+)\oplus(-)$.The corresponding boundary entropy asymptotic expansion is 
\be  \label{hexp1}
s = \frac{1}{2}\ln 2 + \sum_{n=1}^{\infty} \frac{B_{2n}(1-2^{1-2n})}{(2n-1)\alpha^{2n-1} } \, .
\ee

When $\alpha=0$  we have a perturbation by dimension zero operators only and the 
partition function reduces to 
\be
Z_{m_1,m_2} = 2\cosh\left( \pi \sqrt{\nu_{1}+\nu_{2}})\right) \, .
\ee
The boundary entropy is 
\be
s= \ln 2 + \ln[\cosh\left( \pi \sqrt{\nu_{1}+\nu_{2}})\right)] - \pi\sqrt{\nu_1 + \nu_2}\tanh\left( \pi \sqrt{\nu_{1}+\nu_{2}})\right)
\ee
that asymptotically decreases as 
\be
s \approx 2\pi \sqrt{\nu_1+\nu_2} e^{-2\pi  \sqrt{\nu_1+\nu_2} } \, .
\ee
This means that the model approaches a single copy of the free spin boundary condition along an infinitely irrelevant direction. 
This is always the case for boundary perturbations by dimension zero operators.  The corresponding flow can be called component flows as 
the end point is always some component in the original superposition. The approach to the infrared fixed point via an infinitely irrelevant direction is similar to bulk perturbations generating a mass gap for some of the fields.  

The above flows act throughout  a three-dimensional space and it is hard to depict them all on a single diagram. On Figure \ref{diagram1} and Figure \ref{diagram2} we show diagrams representing the $m_1=0$ and $m_2=0$ slices. 
These diagrams include special lines along which the theory flows to intermediate fixed points as well as the asymptotic flow lines that start at those intermediate fixed points and which lie on the boundary of the space of flows from the original UV fixed point.

\begin{center}
\begin{figure}[H]
\centering
\begin{tikzpicture}[scale=1.2]
\begin{scope}[thick, every node/.style={sloped,allow upside down}]
\draw[dashed] (0,3) to[out=0,in=90] node {\midarrow} (3,0); 
\draw[dotted] (0,-3) to[out=0,in=-90] node {\midarrow} (3,0); 
\draw[dotted] (0,-3) to[out=180,in=-90] node {\midarrow} (-3,0);
\draw[dashed] (0,3) to[out=180,in=90] node {\midarrow} (-3,0);
\draw[red] (0,0.2) -- node {\midarrow} (0,3);
\filldraw[white,draw=white] (3,0) circle [x radius=0.25, y radius = 0.25];
\draw (3,0)  node {$\boldsymbol{ (-)}$};
\filldraw[white,draw=white] (-3,0) circle [x radius=0.25, y radius = 0.25];
\draw (-3,0)  node {$\boldsymbol{ (+)}$};

\draw (0,0) node {$\boldsymbol{ (f)\oplus (f)}$};
\filldraw[white,draw=white] (0,3) circle [x radius=0.25, y radius = 0.25];
\draw (0,3) node {$\boldsymbol{ (f)}$};
\filldraw[white,draw=white] (0,-3) circle [x radius=0.75, y radius = 0.75];
\draw (0,-3) node {$\boldsymbol{ (+)\oplus(-)}$};
\draw[blue] (0,-0.2) -- node {\midarrow} (0,-2.8);
\draw (0.75,0) -- node {\midarrow} (2.7,0);
\draw (-0.75,0) -- node {\midarrow} (-2.7,0);
\draw (0.2,0.2) to[out=45,in=130] node {\midarrow} (2.8,0.2);   
\draw (0.15,0.2) to[out=75, in=180]  (1.5, 1.5); 
\draw (1.5,1.5) to[out=0,in=100] node {\midarrow} (2.85,0.2);
\draw (0.1,0.2) to[out=85,in=180] (1,2.3);
\draw (1,2.3) to[out=0,in=99] node {\midarrow} (2.9,0.2);
\draw (-0.2,0.2) to[out=135,in=50] node {\midarrow} (-2.8,0.2);  
\draw (-0.15,0.2) to[out=105, in=0]  (-1.5, 1.5); 
\draw (-1.5,1.5) to[out=180,in=80] node {\midarrow} (-2.85,0.2);
\draw (-0.1,0.2) to[out=95,in=0] (-1,2.3);
\draw (-1,2.3) to[out=180,in=81] node {\midarrow} (-2.9,0.2);
\draw (0.2,-0.2) to[out=-45,in=-130] node {\midarrow} (2.8,-0.2);  
\draw (0.15,-0.2) to[out=-75, in=-180]  (1.5, -1.5); 
\draw (1.5,-1.5) to[out=0,in=-100] node {\midarrow} (2.85,-0.2);
\draw (0.1,-0.2) to[out=-85,in=-180] (1,-2.3);
\draw (1,-2.3) to[out=0,in=-99] node {\midarrow} (2.9,-0.2);
\draw (-0.2,-0.2) to[out=-135,in=-50] node {\midarrow} (-2.8,-0.2);  
\draw (-0.15,-0.2) to[out=-105, in=0]  (-1.5, -1.5); 
\draw (-1.5,-1.5) to[out=-180,in=-80] node {\midarrow} (-2.85,-0.2);
\draw (-0.1,-0.2) to[out=-95,in=0] (-1,-2.3);
\draw (-1,-2.3) to[out=-180,in=-81] node {\midarrow} (-2.9,-0.2);
\end{scope}

\end{tikzpicture}
\caption{The diagram of flows for the $m_{1}=0$ slice. The red flow line (vertical, upwards) depicts  the $h_{-}=0$ flow. 
The blue flow line (vertical, downwards) depicts the $m_{2}=0$  flow. The solid lines correspond to generic $h_{-} m_{2} \ne 0$ flows. The dashed lines depict the asymptotic flows from the fixed point $(f)$ generated by $\sigma_{B}$ while the dotted lines show the asymptotic flows from the fixed point $(+)\oplus(-)$ generated by the identity fields (the component flows). }
 \label{diagram1}
\end{figure}
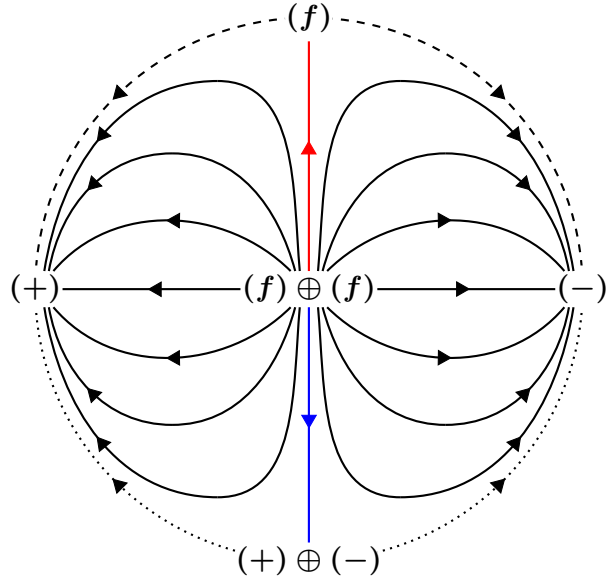
\end{center}


\begin{center}
\begin{figure}[H]
\centering
\begin{tikzpicture}[scale=1.2]
\begin{scope}[thick, every node/.style={sloped,allow upside down}]
\draw[dashed] (0,-3) to[out=0,in=-90] node {\midarrow} (3,0); 
\draw[dashed] (3,0) to[out=90,in=0] node {\midarrow} (0,3); 
\draw[dashed] (0,-3)   to[out=180,in=-90] node {\midarrow} (-3,0)  ;
\draw[dashed] (-3,0) to[out=90,in=180] node {\midarrow} (0,3);
\draw (0,0) node {$\boldsymbol{ (f)\oplus (f)}$};
\filldraw[white,draw=white] (0,3) circle [x radius=0.25, y radius = 0.25];
\draw (0,3) node {$\boldsymbol{ (f)}$};
\filldraw[white,draw=white] (0,-3) circle [x radius=0.75, y radius = 0.75];
\draw (0,-3) node {$\boldsymbol{ (+)\oplus(-)}$};
\draw[blue] (0,-0.2) -- node {\midarrow} (0,-2.8);
\draw (0,0.2) -- node {\midarrow} (0,2.8);
\draw (0.2,0.2) to[out=45,in=-50] node {\midarrow} (0.05,2.8);
\draw (0.75,0) to[out=0,in=-47] node {\midarrow} (0.15,2.8);
\draw (0.2,-0.2) to[out=-60,in=180] node {\midarrow} (1.2,-1.2);
\draw (1.2,-1.2) to[out=0,in=-45] node {\midarrow} (0.2,2.8);
\draw (0.1,-0.2) to[out=-80,in=180] node {\midarrow} (1.3,-2);
\draw (1.3,-2) to[out=0,in=-80] (2.4,0);
\draw (2.4,0) to[out=100,in=-40] node {\midarrow} (0.25,2.8);
\draw (-0.2,0.2) to[out=135,in=230] node {\midarrow} (-0.05,2.8);
\draw (-0.75,0) to[out=180,in=227] node {\midarrow} (-0.15,2.8);
\draw (-0.2,-0.2) to[out=240,in=0] node {\midarrow} (-1.2,-1.2);
\draw (-1.2,-1.2) to[out=180,in=225] node {\midarrow} (-0.2,2.8);
\draw (-0.1,-0.2) to[out=260,in=0] node {\midarrow} (-1.3,-2);
\draw (-1.3,-2) to[out=180,in=260] (-2.4,0);
\draw (-2.4,0) to[out=80,in=220] node {\midarrow} (-0.25,2.8);

\end{scope}

\end{tikzpicture}
\caption{The diagram of flows for the $m_{2}=0$ slice. The blue flow line (vertical, downwards) depicts the $m_{1}=0$  flow.
 The solid lines  correspond to  generic $ m_{1} \ne 0$ flows (that include the vertical upward flow with $h_{-}=0$). The dotted lines show the asymptotic flows from the fixed point $(+)\oplus(-)$ generated by the boundary condition changing  field  $\sigma_{B}^{[+-]}$. }
  \label{diagram2}
\end{figure}
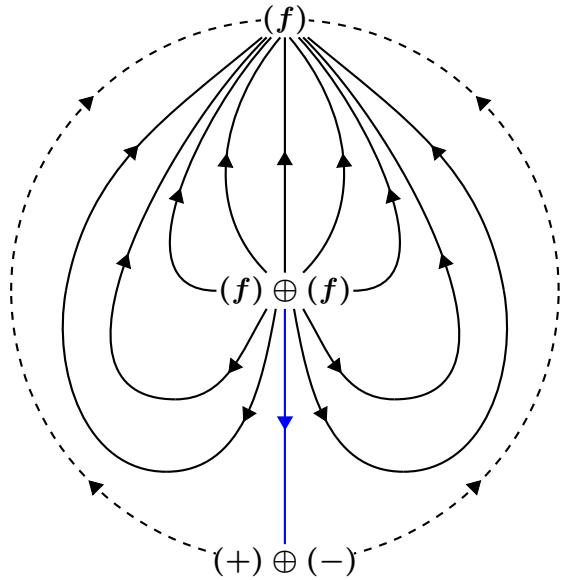
\end{center}


\section{Infrared effective action}\label{effective_sec}
 \setcounter{equation}{0}

Using the asymptotic expansions of the boundary entropy we can also obtain the irrelevant operators along which the flows arrive at the infrared fixed points. Thus, for the pure boundary magnetic field flow ($\nu_1=\nu_2=0$) we get from (\ref{hexp1})
\be \label{sash}
s = \frac{1}{2}\ln 2 +\frac{1}{12\alpha} - \frac{7}{720 \alpha^3} + \dots 
\ee
As the leading correction in this expansion scales as $1/\beta$ this means that the leading irrelevant perturbation near the infrared fixed point has dimension 2 and thus must be proportional to the stress-energy tensor $T$. Comparing the expectation value 
\be
\langle T\rangle = -\frac{\pi^2}{12 \beta^2} 
\ee
with the coefficient in (\ref{sash}) we see that the leading term in the infrared effective action is 
\be \label{Seff_lead}
S_{\rm IR} = \frac{1}{4\pi^2 h^2_{-}} \int\limits_{0}^{\beta}\!\! d\tau\,  T(\tau) + \dots 
\ee
The same leading term was obtained in  \cite{Toth}, \cite{me} by other methods. Such a perturbation would normally be expected 
to contribute to the free energy (and thus to the boundary entropy) at higher orders in the effective coupling via 
the integrated connected $n$-point functions. For a perturbing operator ${\cal O}$ of dimension $\Delta$ the second order correction 
is proportional to the integral 
\be \label{quadratic_int}
\int\limits_{0}^{\beta}\!\! d\tau_{1} \int\limits_{0}^{\beta}\!\! d\tau_{2} \langle {\cal O}(\tau_{1}) {\cal O}(\tau_{2})\rangle_{c;\beta} 
= (2\beta)^{2(1-\Delta)}\frac{\Gamma(1/2-\Delta) }{4\sqrt{\pi}\Gamma(1-\Delta) }
\ee
defined via analytic continuation in dimension (see e.g.  \cite{me_pert}). Here $\langle \dots \rangle_{c;\beta}$ stands for 
a connected correlator on a cylinder of circumference $\beta$. 
The integral (\ref{quadratic_int}) vanishes for any positive integer valued $\Delta$. This explains the absence of a term of order $1/\alpha^2$ 
in (\ref{sash}). Furthermore, an integrated correlator of three primary fields is given by \cite{me_pert} 
\bea \label{cubic_int}
&&\int\limits_{0}^{\beta}\!\! d\tau_{1} \int\limits_{0}^{\beta}\!\! d\tau_{2}\int\limits_{0}^{\beta}\!\! d\tau_{3} 
\langle {\cal O}_{1}(\tau_{1}) {\cal O}_{2}(\tau_{2}){\cal O}_{3}(\tau_{3})\rangle_{c;\beta} 
= \frac{1}{2}(C_{123}+C_{213})(2\beta)^{3-\Delta_{1}-\Delta_{2}-\Delta_{3}} 
 \nonumber \\
&& \times \frac{\Gamma\left(\frac{2-\Delta_{1}-\Delta_{2}-\Delta_{3}}{2} \right)\Gamma\left( \frac{1+\Delta_{1}+\Delta_{2}-\Delta_{3}}{2} \right)\Gamma\left( \frac{1+\Delta_{1}+\Delta_{3}-\Delta_{2}}{2} \right)\Gamma\left( \frac{1+\Delta_{3}+\Delta_{2}-\Delta_{1}}{2} \right)}{(4\pi)^{3/2}\Gamma(1-\Delta_{1})\Gamma(1-\Delta_{2})\Gamma(1-\Delta_{3})} \, .
\eea
We see that in the absence of a resonance $1+ \Delta_{i} + \Delta_{j} = \Delta_{k}$ this quantity vanishes when at least one of the fields ${\cal O}_{i}$ has an integer dimension. In particular this implies that for the perturbation (\ref{Seff_lead}) there is no contribution 
of order $1/\alpha^3$ to the boundary entropy. (Although $T$ is only a quasiprimary it mixes under general conformal transformations only with the identity field that drops out in the connected correlators.) This means that the third term in the expansion (\ref{sash})  must come from an additional perturbation by an operator of dimension 4. The only operator available is the fermion bilinear
\be
T_{4}(\tau)  \equiv :\!\psi \partial_{z}^{3} \psi \!:\Bigr|_{z=i\tau}
\ee 
for which the expectation value on a cylinder is 
\be
\langle T_{4} \rangle = -\frac{7}{60} \left( \frac{\pi}{\beta}\right)^4 \, .
\ee
This implies that the next to leading term in the effective action is 
\be \label{Seff2}
S_{\rm IR} = \frac{1}{4\pi^2 h^2_{-}} \int\limits_{0}^{\beta}\!\! d\tau\,  T(\tau) -  
\frac{1}{24 (2\pi)^4  h^6_{-}} \int\limits_{0}^{\beta}\!\! d\tau\,  T_{4} (\tau) + \dots 
\ee
Using formulae\footnote{The connected three point function $\langle T\, T\, T_{4}\rangle$ on the boundary of a half-cylinder can be obtained via a mapping  to the plane. The operator $T_{4}$ is represented on the plane by a combination containing the identity, $T$ and $\partial T$. Nevertheless the desired three-point function  can  be represented as a linear combination of 3-point functions of the type standing in the integrand of  (\ref{cubic_int}).} (\ref{quadratic_int}),  (\ref{cubic_int}) we can show that the two terms in (\ref{Seff2}) do not generate a 
correction of order $1/\alpha^5$ to the boundary entropy and hence a new term with dimension 6 
operator has to be added which has to be\footnote{Note that the lowest dimension operator quartic in the free fermion field is 
$:\! \psi \partial_{\tau}\psi \partial^{2}_{\tau}\psi\partial^{3}_{\tau}\psi\!:$ that has dimension 8.} proportional to $T_{6} = :\!\psi \partial_{z}^{5} \psi\!:$. 
Given that the theory at hand is Gaussian it is natural to assume that all higher order operators in 
the effective action are fermion bilinears 
\be
T_{2n}(\tau)  \equiv :\!\psi \partial_{z}^{2n-1} \psi \!:\Bigr|_{z=i\tau}
\ee
and that the complete effective action has the form 
\be \label{Seff_complete} 
S_{\rm IR} = \lambda_{2}  \int\limits_{0}^{\beta}\!\! d\tau\,  T(\tau) + \sum_{n=2}^{\infty} 
 \lambda_{2n} \int\limits_{0}^{\beta}\!\! d\tau\,  T_{2n} (\tau)
\ee
where $\lambda_{2n}=\lambda_{2n}(h_{-})$ are the effective couplings with 
\be 
\lambda_{2} =  \frac{1}{4\pi^2 h^2_{-}} \, , \qquad \lambda_{4}= - \frac{1}{24 (2\pi)^4  h^6_{-}} \, .
\ee
The expectation values of the fermion bilinears  on a cylinder can be obtained in a closed form using point splitting 
regularisation 
\be \label{Tvevs}
\langle T_{2n} \rangle = -\left(\frac{\pi}{\beta}\right)^{2n}\lim_{z\to 0} \frac{\partial^{2n-1}}{\partial z^{2n-1}} \left( 
\frac{1}{\sinh(z)} - \frac{1}{z} \right) = \left(\frac{\pi}{\beta}\right)^{2n} \frac{(2^{2n-1} -1)}{n} B_{2n} \, .
\ee
The same answer was obtained by other methods in \cite{FS}, \cite{DW}.
To make a complete analysis of the infrared effective action we need to have the integrals of higher order 
correlators under control. We present now a general 
 argument that shows that all such integrals 
vanish. The integrands in the integrals of interest are connected correlation functions  of the operators $T$ and $T_{2n}$, 
$n\ge 2$ taken on the cylinder and restricted to a circle. They are thus restrictions of meromorphic  functions defined on the entire cylinder. Using this we regularise the integrals by shifting the integration contours away from this circle for example by 
considering\footnote{This regularisation was considered in \cite{RS}, \cite{RS2} for perturbations by marginal operators. One can make it more symmetric by symmetrising over permutations of the operator insertions as well as over $\epsilon \to -\epsilon$.} 
\be
\int\limits_{0}^{\beta}\!\! d\tau_{1} \dots \int\limits_{0}^{\beta}\!\! d\tau_{k} \langle T_{2n_1}(\tau_{1}) T_{2n_2} (\tau_{2}-i\epsilon) T_{2n_3} (\tau_{3}-2i\epsilon) \dots 
	T_{n_{k}}(\tau_{k}-(k-1)i\epsilon) \rangle_{c; \beta} \, .
\ee
Clearly in the limit $\epsilon \to 0$ we recover the original integrand. On the other hand the outer contour of integration parameterised by $\tau_{k}$ can now be deformed to infinity. Since at infinity the connected correlation function decays exponentially the integral vanishes. 
This extends to all orders the vanishing of the second and cubic order corrections observed above.  
Therefore the boundary entropy corresponding to (\ref{Seff_complete} ) can be obtained entirely from the 
expectation values (\ref{Tvevs}). Matching the resulting expression with the complete expansion  (\ref{hexp1})
we obtain 
\be \label{lambda_2n}
\lambda_{2n} = -\frac{1}{(2h_{-}^2)^{2n-1}(2\pi)^{2n}(2n-1)} \, , \quad n\ge 2 \, . 
\ee

The same kind of analysis of the infrared effective action can be repeated in the presence of other couplings. 
Thus, in the generic case $\alpha \nu_{2} \ne 0$ the two leading terms in the expansion of $s$ are 
\be
s = -\frac{1}{2}\ln 2 + \frac{\nu_1 + \nu_2}{12\alpha \nu_{2}} + 7\frac{3\nu_1 \nu_2 \alpha^2 - (\nu_1+\nu_2)^3}{720\alpha^3 \nu_2^3} + \dots 
\ee
that gives the first two couplings  in the effective action (\ref{Seff_complete} ) 
\be
\lambda_{2} = \frac{m_{1}^2 + m_{2}^2}{(2\pi)^2 h_{-}^2 m_{2}^2} \, , \quad 
\lambda_{4} = \frac{1}{48\pi^4} \left(  \frac{6\pi m_{1}^2}{h_{-}^2 m_{2}^4} - \frac{1}{8h_{-}^6} 
\left(1+ \frac{m_{1}^2}{m_2^2}\right)^2 \right) \, . 
\ee
We clearly see that this couplings are singular when $\alpha \nu_{2}$ tends to zero. This reflects the obvious fact that the $\beta \to \infty$ limit does not commute with the $\alpha \nu_{2} \to 0$ limit in the coupling space. 
It would be interesting to investigate further this singularity in particular in conjunction with the RG operators discussed in 
\cite{me_last}. It is straightforward to obtain all other couplings in the effective action. They can be expressed in terms of 
symmetrised  sums  of odd inverse powers of the roots $k_{i}$. We discuss the effective action (\ref{Seff_complete}) a bit more in the last section of the paper.

\section{Concluding remarks} \label{conclude_sec}
In this section we would like to point some directions in which the present work can be extended. 
Firstly we would like to note that the three-coupling model (\ref{S2}) with $h_{+}=0$ can be further investigated. It would be interesting to calculate some local correlation functions, in particular the local magnetisation that would further elucidate the boundary bound states and the phase structure of the model. Secondly, this model presents a nice toy model of RG flows in a multi-coupling space with several fixed points present. It would be interesting to investigate the singularities in the space of flows further by finding the RG operators and investigating the transport of local operators along the flows as discussed in   \cite{me_last}. The complete infrared effective action (\ref{Seff_complete}) 
that we obtained should be a useful tool here. It is interesting to note its much greater simplicity in comparison to the infrared expansion obtained in \cite{me}  using the mode truncation regularisation and the Schriffer-Wolf method. 

Furthermore, there are several directions in which the present work could go beyond the three-coupling Gaussian model. 
Integrable boundary conditions in the presence of bulk four-fermion interactions were found in \cite{4fermi} at the classical level. 
It is foreseeable then that for some restricted values of the couplings the four-coupling model (\ref{S2}) is integrable in the presence of 
the boundary four-fermion interaction governed by $h_{+}$. 
To study this question systematically  we searched for classically integrable boundary conditions in the model  (\ref{S2}).
We found that the dimension 4 current $(T_{4}, \bar T_{4})$ and the dimension 6 current $(T_{6}, \bar T_{6})$ are each 
 classically conserved if and only if $m_{1}h_{+}h_{-}=0$. Some details of these calculations are presented in the appendix. This suggests that if the discontinuities are present at all the four coupling model is integrable only if $h_{1}=\pm h_{2}$.  It may be possible to analyse the generic $N=2$ model approximately using the Hubbard-Stratonovich transformation. We plan to pursue this direction in future work. 

For $N\ge 3$ we can represent the boundary perturbation (\ref{bmf_N}) in terms of a larger number of boundary fermions similar to 
(\ref{S2}). The quadratic models do not seem to contain any new interesting cases.  It would be interesting to extend the integrability 
analysis done in the appendix to this case to see whether there are any new integrable models with non-trivial four-fermion boundary interaction.  

Although we relied heavily on the free fermion representation we find our results suggestive that there may be interesting integrable boundary theories which are obtained by deforming a superposition of boundary conditions in a minimal model by a collection of boundary condition changing operators with weights zero and $h_{(1,3)}$ 
that are known to give integrable deformations for irreducible Cardy boundary conditions. As the exact partition function (\ref{Z3ren}) we obtained in the present model satisfies non-trivial factorisation conditions, one way to search for new integrable deformations could be using the results of \cite{Ingo_T} and generalising them to deformations in superpositions. 

One of the motivations for the present work was the problem of obtaining some description of a complete set of boundary RG flows in the Ising model. 
This space includes all flows generated by relevant boundary fields starting from an arbitrary superposition of the free and fixed spin conformal boundary conditions.  Unlike any space of bulk RG flows, in which there always could be flows from an arbitrary UV fixed point  
 with a high enough central charge ending in a given IR CFT, a space of boundary flows  in a fixed bulk CFT is much more contained. The bulk CFT does not change and all conformal boundary conditions may be  known as is the case in the Ising model or more generally in any Virasoro minimal model. With these observations in mind we could try a bottom-up approach to describing the space of such flows.
 An irrelevant boundary deformation in  a theory where all conformal boundary conditions are known, in the deep UV region would  either lead to a known conformal boundary condition or not to lead to any local boundary condition at all. Starting with \cite{SZ} an investigation into the structure of the space of $T\bar T$ and more general irrelevant integrable deformations has been going on. In   \cite{LeClair1}, 
 \cite{LeClair2} the question of finding  additions of higher dimension  irrelevant operators that ensure a UV completion  was considered. A similar question can be posed for boundary deformations -- find integrable deformations by irrelevant operators that have a UV completion. Note that the boundary  deformation by the stress-energy operator $T$ has many similarities  to the $T\bar T$  bulk deformation. As the stress-tensor operator is universal the deformation can be unambiguously defined at any value of the coupling. 
 It is in some sense topological as the perturbation contour of integration can be moved away from the boundary using holomorphy. 
 The perturbation's behaviour depends on the sign of the coupling with one direction being singular. This singularity is signified by the divergence of a cylinder partition function\footnote{This singularity of the boundary $T$-perturbation has been discussed in detail in \cite{me_Dermot}.} at a critical cylinder length that is similar to the Hagedorn behaviour of the partition function in the $T\bar T$ deformation case. The effective action (\ref{Seff_complete}) with the couplings (\ref{lambda_2n}) give an explicit example of how adding an infinite sequence of  irrelevant boundary interactions to a $T$-deformation leads to a UV-complete theory.
 The perturbing terms in (\ref{Seff_complete}) are the KdV charges and the cylinder partition function can be built from the chiral traces which were recently considered in \cite{DW} in connection with generalised Gibbs ensembles. 
 We believe this connection needs to be investigated further and may lead to interesting results on he space of all boundary flows in the Ising model.  We leave this to future work.

 \setcounter{equation}{0}

\begin{center}
{\bf \large Acknowledgements} 
\end{center}
The author thanks R. Weston  for useful discussions and the anonymous referees for valuable comments on the first version of the paper. 

\appendix
\renewcommand{\theequation}{\Alph{section}.\arabic{equation}}
\setcounter{equation}{0}

\section{Classical integrability of the generic four coupling model}
\setcounter{equation}{0}
In this appendix we investigate classical integrability of the theory defined in (\ref{S2}). For simplicity we consider the conformal point $m=0$. The boundary conditions in the generic  $h_{+}h_{-}\ne 0$ case read 
\bea \label{bcs_gen}
 \psi - \bar \psi &=& 4\pi h_{-} a  - 4\pi i h_{+} abc\, ,  \nonumber \\ 
 \dot a &=& -ih_{-}(\psi + \bar \psi) -h_{+}bc(\psi + \bar \psi) - m_{1}ib   \, ,  \nonumber \\  
 \dot b &=& h_{+}ac(\psi + \bar \psi)+im_{1}a - im_{2}c  \, ,  \nonumber \\ 
 \dot c &=& -h_{+}ab(\psi + \bar \psi)+im_{2}b \,   
\eea
where for brevity we suppress the argument $t$ in all functions.

The bulk currents whose conservation we investigate are 
\be
T(z) = -\frac{1}{2}:\!\psi\partial_{z}\psi\!: \, , \qquad T_{2n}(z)  = :\!\psi \partial^{2n-1}_{z}\psi\!: \, , \enspace n\ge 2
\ee
and their anti-holomorphic counterparts in which $\partial_{z}$ is replaced by $ \partial_{\bar z}$. For the stress energy tensor 
the boundary conservation equation on a half plane is 
\be \label{energy_cons}
(T - \bar T)\Bigr|_{z=\bar z=t} = -2\pi i \partial_{t} \theta(t)
\ee
where $\theta(t)$ is the boundary part of the stress-energy tensor in standard normalisation (see e.g. \cite{gThm2} ). A higher spin current $T_{2n}$ is conserved if 
\be
(T_{2n} - \bar T_{2n})\Bigr|_{z=\bar z=t} =  \partial_{t} \theta_{2n}(t) 
\ee
for some boundary operator $\theta_{2n}(t)$. As a warm up we first derive $\theta$ and then investigate the conservation of 
$T_{4}$ and $T_{6}$. We will be using the classical boundary conditions (\ref{bcs}) in which all fermionic fields are considered to be anti-commuting.

We start by noting the following expressions for derivatives of dimension zero operators that follows from (\ref{bcs_gen}) 
\bea \label{id2} 
\pt (ab) &=& ih_{-}b(\psi + \bar \psi) - im_{2}ac \, , \nonumber \\
\pt(ac) & = & ih_{-}c(\psi + \bar \psi) - im_{1}bc + im_{2}ab \, , \nonumber \\
\pt(bc) &=& im_1 ac \, , \nonumber \\
\pt(abc) &= & -ih_{-} bc(\psi + \bar \psi) \, .
\eea

To find $\theta$ we  differentiate the first equation in (\ref{bcs_gen}) to obtain
\be \label{eq1}
\pt\psi - \pt\bar \psi=-i4\pi h_{-}^2(\psi + \bar \psi) -8\pi h_{-}h_{+}bc (\psi + \bar \psi) -4\pi h_{-}m_{1}ib \, .
\ee 
Multiplying this equation by $\psi$ and separately by $\bar \psi$ on the left we obtain the following two equations 
\bea
\psi\pt \psi - \psi\pt\bar \psi &=& -i4\pi h_{-}^2 \psi \bar \psi - 8\pi h_{-}h_{+}\psi\bar\psi bc -4\pi h_{-}m_{1}i\psi b \, , \nonumber \\
\bar \psi\pt \psi - \bar \psi\pt\bar \psi &=&-i4\pi h_{-}^2 \bar \psi  \psi -8\pi h_{-}h_{+}\bar\psi\psi bc -4\pi h_{-}m_{1}i\bar\psi b
\eea 
Adding these two equations we obtain 
\be \label{int1}
\psi\pt \psi - \bar \psi \pt \bar \psi = \pt(\psi\bar \psi ) - 4\pi h_{-}m_{1}i(\psi + \bar \psi)b \, .
\ee
From (\ref{id2}) we obtain 
\be \label{td}
m_{1}\pt(ba) + m_{2}\pt(cb) = ih_{-}m_{1}(\psi + \bar \psi) b \, .
\ee
The last formula allows us to rewrite (\ref{int1}) as 
\be
\psi\pt \psi - \bar \psi \pt \bar \psi =  4\pi\pt\Bigl[ \frac{1}{4\pi} \psi\bar \psi  +  m_{1}ab + m_{2}bc \Bigr] \, .
\ee
Comparing this with (\ref{energy_cons}) we find that 
\be
\theta(t) = -i\Bigl[ \frac{1}{4\pi} \psi\bar \psi  +  m_{1}ab + m_{2}bc \Bigr]  
\ee
that can be also rewritten as 
\be
\theta(t) = -\Bigl[ \frac{ih_{-}}{2}a(\psi+ \bar \psi) + \frac{h_{+}}{2}abc(\psi + \bar \psi)  +  im_{1}ab + im_{2}bc \Bigr]
\ee
that has the standard form of the sum of perturbing operators multiplied by their beta functions.

We next take up the dimension 4 currents: $T_{4}$ and $\bar T_{4}$. To find their difference on the boundary we 
differentiate (\ref{eq1}). From that, using (\ref{bcs_gen}) and (\ref{id2}),  we obtain 
\bea
\pt^2\psi - \pt^2\bar \psi &&= -4\pi ih_{-}^2(\pt \psi + \pt\bar\psi) -8\pi h_{-}h_{+} bc (\pt \psi + \pt\bar\psi) \nonumber \\
&& + 24\pi i h_{-}h_{+} ca (\psi + \bar \psi) + 4\pi m_{1}^2 h_{-} a - 4\pi m_{1}m_{2}h_{-} c
\eea 
Differentiating one more time we get 
\bea
&& \pt^3\psi - \pt^3\bar \psi = -4\pi i h_{-}^2(\pt^2\psi + \pt^2\bar\psi) -8\pi h_{-}h_{+} bc(\pt^2 \psi + \pt^2\bar\psi)  \nonumber \\
&& +20\pi h_{-}h_{+}m_{1}ica(\pt\psi + \pt\bar\psi) + 16\pi m_{1}^2h_{-}h_{+} cb(\psi + \bar \psi)   \nonumber \\
&& + 16\pi m_{1}m_{2}h_{-}h_{+} ab(\psi + \bar \psi)  -4\pi m_{1}^2 h_{-}^2 i(\psi + \bar \psi)  
-4\pi m_{1} h_{-}(m_{1}^2 + m_{2}^2) ib
\eea 
Next we multiply both sides of this equation by $\psi + \bar \psi$ from the left to obtain 
 \bea \label{T4diff}
 &&(T_{4} - \bar T_{4}))\Bigr|_{z=\bar z=t}=\psi\pt^3\psi - \bar\psi\pt^3\bar\psi = X_{1} -4\pi i h_{-}^2 X_{2} -8\pi h_{-}h_{+} bc X_2  \nonumber \\
 && + 20 \pi h_{-}h_{+}m_{1} ica X_{3} -4\pi m_{1}h_{-}(m_1^2 + m_2^2)i(\psi + \bar \psi) b 
 \eea 
 where 
 \be
 X_{1} \equiv  \psi \pt^3 \bar \psi - \bar \psi \pt^3\psi = \pt [(\pt \bar \psi )\pt \psi + \psi \pt^2 \bar \psi - \bar \psi\pt^2\psi ] \, ,
 \ee
 \be
 X_{2} \equiv (\psi + \bar \psi)(\pt^2 \psi + \pt^2 \bar \psi) = \pt[ \psi\pt\psi + \bar\psi\pt\bar\psi + \psi\pt\bar\psi + \bar\psi\pt\psi]
 \ee
 \be 
 X_{3} \equiv (\psi + \bar \psi)(\pt \psi + \pt \bar \psi) \, .
 \ee
 Note that 
 \be
 X_{2} = \pt X_{3} \, .
 \ee
 
 Since by (\ref{td}) the term containing $(\psi + \bar \psi) b $ is also a derivative the problematic terms are 
 \be
 \Xi \equiv -8\pi h_{-}h_{+} bc X_2 + 20 \pi h_{-}h_{+}m_{1} ica X_{3}  \, .
 \ee 
 We can rewrite this as 
 \bea
&&  \Xi = -8\pi h_{-}h_{+} bc \pt X_3 + 40 \pi h_{-}h_{+}m_{1} ica X_{3}   \nonumber \\
&& = \pt( -8\pi h_{-}h_{+} bc  X_3 ) + 8\pi  h_{-}h_{+} \pt(bc) X_{3} + 20 \pi h_{-}h_{+}m_{1} ica X_{3} \nonumber \\
&&  = \pt( -8\pi h_{-}h_{+} bc  X_3 )   + (20 \pi - 8\pi) h_{-}h_{+}m_{1} ica X_{3} \nonumber \\
&& = \pt( -8\pi h_{-}h_{+} bc  X_3 )   + 12\pi h_{-}h_{+}m_{1} ica X_{3}
 \eea 
 and the  obstruction for the difference $T_{4}-\bar T_{4}$ to be a derivative now looks very compact. If $ h_{-}h_{+}m_{1} \ne 0$ then to show that the dimension 4 current is not conserved we have  to prove that
 \be  \label{oop}
 {\cal O} \equiv ca X_{3} = ca(\psi + \bar \psi)(\pt \psi + \pt \bar \psi) = 4ca \psi \pt \psi  - 8\pi i m_{1}h_{-} abc \psi
 \ee 
 cannot be written as a derivative of another operator. Although applying the derivative always rises the dimension by one we cannot use the dimension grading as 
 the couplings also carry the natural dimension and there could be divisions by the couplings present. Instead we can rely on the grading provided by the number of $\psi$-fields and their derivatives present in a composite operator. As no derivative of $\psi$ or $\bar \psi$ is present 
 in the boundary conditions (\ref{bcs_gen}) any operator can be represented as a finite linear combination of operators of the form 
 \be \label{basis_op}
 f(a,b,c)\pt^{k_{1}}\psi \dots  \pt^{k_{n}}\psi \, , \enspace k_{i} \ge 0
 \ee
 where $f(a,b,c)$ is a polynomial in boundary fermions. The derivative of such an operator is a linear combination of operators 
 of the same form. Schematically we have 
 \bea \label{schem}
 &&\partial_{t}  [f(a,b,c)\pt^{k_{1}}\psi \dots  \pt^{k_{n}}\psi ] = f_{1}(a,b,c) \pt^{k_{1}}\psi \dots  \pt^{k_{n}}\psi  
 \nonumber \\
 && + f_{2}(a,b,c) \psi \pt^{k_{1}}\psi \dots  \pt^{k_{n}}\psi  + f(a,b,c) \pt (\pt^{k_{1}}\psi \dots  \pt^{k_{n}}\psi )
 \eea
 where $f_{1}$ and $f_{2}$ are some new polynomials.
 This means that when we take the derivative of (\ref{basis_op}) we obtain only operators with either $n$ or $n+1$ $\psi$-fields and the total number of derivatives acting on them can increase only by one. This immediately implies that 
 there are only 4 possible operator types  whose derivative could give  
 ${\cal O}$. Schematically we can write them as 
 \bea
&&    P = p(a,b,c) \psi \pt \psi \, , \qquad Q=q(a,b,c) \pt \psi \, \nonumber \\
&&   R=r(a,b,c) \psi \, , \qquad S=s(a,b,c) 
 \eea
 where $p,q,r,s$ are some polynomials in boundary fermions whose coefficients can depend on the couplings . 
  In the notation introduced in (\ref{schem}) the derivatives of these operators can be written as 
 \be
 \pt P = p_1(a,b,c)\psi\pt \psi + p(a,b,c)\psi\pt^2\psi  \, , 
 \ee
  \be
 \pt Q = q_1(a,b,c)\pt \psi + q_2(a,b,c) \psi \pt \psi + q(a,b,c)\pt^2 \psi  \, , 
 \ee
 \be
 \pt R =r_1(a,b,c)\psi + r(a,b,c) \pt \psi \, , 
 \ee
 \be
 \pt S = s_1(a,b,c) + s_2(a,b,c)\psi  \, . 
 \ee
 From these expressions we see that the only way to ensure $\psi\pt \psi$ is present is to include at least one of the operators $P$ and $Q$. But the first one necessarily generates $\psi\pt^2\psi$ while the second generates $\pt^2 \psi$ which are not present in ${\cal O}$. 
We conclude that the operator ${\cal O}$ cannot be written as a derivative and the current $(T_{4}, \bar T_{4})$ is conserved if and only if $h_{-}h_{+}m_1 =0$. 

In a similar fashion we also analysed the conservation of the current $(T_{6}, \bar T_{6})$. Omitting the details, we found that 
\bea \label{T6diff}
(T_{6} - \bar T_{6})\Bigr|_{z=\bar z=t} = && 20\pi i m_{1}h_{+}h_{-}[(m_1^2 + m_2^2) {\cal O}  +4ac\pt \psi \pt^2\psi + 16\pi i h_{-}^2 ac \psi \pt^2\psi 
 \nonumber \\
&& -8\pi i m_1 h_{-}abc \pt^2 \psi] + \pt Y 
\eea
where $Y$ is some local operator and ${\cal O}$ is given in (\ref{oop}). Analysing the expression in the square brackets as above 
we find that it cannot be represented as a derivative. Hence the current $(T_{6}, \bar T_{6})$ is conserved if and only if $h_{-}h_{+}m_1 =0$.
Moreover, the conclusion holds even if we take a linear combination of $T_{4}$ and $T_{6}$ which removes the ${\cal O}$ field from (\ref{T6diff}). The essential obstruction in (\ref{T4diff}) and (\ref{T6diff}) for being represented as derivatives is the presence of operators $ac\psi\pt \psi$ and $ac\psi\pt^3 \psi$ which are the operators with the highest number of derivatives in each expression.

\end{document}